\documentclass[conference]{IEEEtran}
\IEEEoverridecommandlockouts
\usepackage{cite}
\usepackage{amsmath,amssymb,amsfonts}
\usepackage{algorithmic}
\usepackage{graphicx}
\usepackage{textcomp}
\usepackage{xcolor}
\usepackage[ruled,linesnumbered,lined]{algorithm2e}
\input{Qcircuit}

\begin{document}

\title{Quantum One-class Classification With a Distance-based Classifier
\thanks{This work is supported by research grants from CNPq, CAPES and FACEPE (Brazilian research agencies).}
}

\author{\IEEEauthorblockN{Nicolas M. de Oliveira, Lucas P. de Albuquerque}
\IEEEauthorblockA{\textit{Centro de Informática} \\
\textit{Universidade Federal de Pernambuco}\\
Recife, Brazil \\
\{nmo, lpa\}@cin.ufpe.br} \and
\IEEEauthorblockN{Wilson R. de Oliveira}
\IEEEauthorblockA{\textit{Departamento de Estatística e Informática} \\
\textit{Universidade Federal Rural de Pernambuco}\\
Recife, Brazil \\
wilson.rosa@gmail.com}
\and \hspace{6.25cm}
\IEEEauthorblockN{\hfill Teresa B. Ludermir, Adenilton J. da Silva}
 \IEEEauthorblockA{\hspace{6.25cm}\textit{Centro de Informática} \\
\hspace{6.25cm} \textit{Universidade Federal de Pernambuco}\\
\hspace{6.25cm} Recife, Brazil \\
\hspace{6.25cm} \{tbl, ajsilva\}@cin.ufpe.br}
}

\maketitle

\begin{abstract}
The advancement of technology in Quantum Computing has brought possibilities for the execution of algorithms in real quantum devices. 
However, the existing errors in the current quantum hardware and the low number of available qubits make it necessary to use solutions that use fewer qubits and fewer operations, mitigating such obstacles. Hadamard Classifier (HC) is a distance-based quantum machine learning model for pattern recognition. We present a new classifier based on HC named Quantum One-class Classifier (QOCC) that consists of a minimal quantum machine learning model with fewer operations and qubits, thus being able to mitigate errors from NISQ (Noisy Intermediate-Scale Quantum) computers. Experimental results were obtained by running the proposed classifier on a quantum device and show that QOCC has advantages over HC.
\end{abstract}


\section{Introduction}\label{sec:intro}

Quantum Computing (QC)~\cite{nielsen2010quantum} can solve some problems more efficiently than any known classical algorithm. Examples of quantum speedup are Shor's factoring algorithm~\cite{shor1994} and Grover's search algorithm~\cite{grover1996}. Characteristics such as quantum parallelism and other phenomena only observed in quantum mechanics increased research interest of QC to problems without known efficient algorithmic solutions.

Quantum Machine Learning (QML)~\cite{biamonte2017quantum} is an area of quantum computing that combines artificial intelligence techniques with the power of quantum computing. 
Several works propose Quantum Machine Learning models and quantum supervised learning problems. More related to our proposal, there are the works presented in~\cite{blank2020quantum,park2020theory,kathuria2020implementation,ruan2017quantum,schuld2017implementing,neumann2020classification}. Such works introduce distance-based quantum classifiers with similarities to the classifier proposed in this paper and will be described in Section~\ref{sec:qml}.

We investigate the Hadamard Classifier (HC)~\cite{schuld2017implementing} and present an improved minimal classifier named Quantum One-Class Classifier (QOCC) that aims to mitigate errors from NISQ (Noisy Intermediate-Scale Quantum) devices \cite{Preskill2018quantumcomputingin}. QOCC is based on HC and resembles the operation of a \textit{Probabilistic Quantum Memory} (PQM)~\cite{trugenberger2001probabilistic}. We reduced HC by removing its class and index quantum registers maintaining accuracy. Besides that, HC needs a 2-qubit measurement to perform the classification, and QOCC requires only a one-qubit measurement. This modification results in a reduction in measurement errors from current noisy quantum devices. Both HC and QOCC are quantum machine learning models that use quantum interference to classify new input data.

We perform experiments on IBM Quantum Experience~\cite{ibmq} to validate our classifier, the first experiment in an error-free simulation environment, and then in two non-error-corrected quantum processors~\textit{ibmq\_athens} and~\textit{ibmq\_santiago}. In the error-free experiment, results show that QOCC accuracy is equivalent to HC accuracy. QOCC has equivalent or better accuracy in real quantum devices. We present the QOCC as a minimum classifier that has competitive results compared to HC (even with fewer operations and qubits) and classical classifiers. Also, we provide an update on HC performance in current quantum devices. 

The remainder of this paper is divided into 6 sections. Section~\ref{sec:qc} summarizes the basic principles of quantum computing. Section~\ref{sec:qml} describes some quantum machine learning models related to this work. Section~\ref{sec:hc} describes the Hadamard Classifier that motivated our proposal. Section~\ref{sec:qocc} presents the main results of this work: a description of the Quantum One-Class Classifier. Section~\ref{sec:results} gives details of the experiments, results, and a discussion. Finally, Section~\ref{sec:conclusion} is the conclusion.

\section{Quantum Computing}\label{sec:qc}

A quantum computer is a machine capable of performing computational calculations and operations based on inherently quantum properties. Analogous to the classic bit, the unit of quantum information is the quantum bit, or \textit{qubit}. The logical values ``0'', ``1'', or any superposition of these can be assigned to a qubit. This superposition consists of a linear combination of the states of the computational basis described by a vector as described in Eq. \eqref{eq:psi}, where $\alpha$ and $\beta$ are probability amplitudes associated with the respective states and $|\alpha|^{2}+|\beta|^{2}=1$.

\begin{equation}\label{eq:psi}
    |\psi\rangle=\alpha|0\rangle+\beta|1\rangle=\begin{bmatrix}
    \alpha\\ 
    \beta
    \end{bmatrix}
\end{equation}

One of the main characteristics of quantum computing compared to classical computing is the superposition of states. This superposition allows quantum computing to obtain a high degree of computational parallelism. With $n$ qubits, we can create the superposition described in Eq. \eqref{eq:parallel}, where $\alpha_{i}$ are probability amplitudes associated with $i$ states. Thus, $n$ qubits can represent $2^{n}$ combinations of states.

\begin{equation}\label{eq:parallel}
    |\psi\rangle:=\sum_{i=0}^{2^{n}-1}\alpha_{i}|i\rangle
\end{equation}

Operations under a quantum state are performed by unitary operators. Given a $U:\mathbb{V}\rightarrow \mathbb{V}$ operator, with $\mathcal{V}$ denoting a vector space, $U$ is said to be unitary when its inverse is equal to its conjugate transpose. That is, $UU^{\dagger} = U^{\dagger}U = I$, where $I$ designates the identity operator. 
A quantum operator acts linearly on vectors (Eq.~\eqref{eq:operator}).

\begin{equation}\label{eq:operator}
    U|\psi\rangle=U\left ( \sum_{i=0}^{2^{n}-1}\alpha_{i}|i\rangle \right ) = \sum_{i=0}^{2^{n}-1}\alpha_{i}U|i\rangle
\end{equation}

Quantum measurements are performed by operators who act on the quantum state to determine the result of the computation that was performed. Given the state in Eq.~\eqref{eq:psi}, one can use the measurement operators $M_0=|0\rangle\langle 0|$ and $M_1=|1\rangle\langle 1|$ to obtain the probability that the measurement result is 0 ($|\alpha|^2$) or 1 ($|\beta|^2$), respectively.

Quantum circuits are one of the ways available to represent quantum computing. It is the quantum circuits that determine which and in what order the operators are applied to one or more qubits. Examples of quantum operators/gates and their respective actions are: \textit{Not} \eqref{eq:not}, \textit{Hadamard} \eqref{eq:hadamard}, \textit{R$_y$} \eqref{eq:ry} and \textit{Controlled-not (CNOT)} \eqref{eq:cnot}.

\begin{equation}\label{eq:not}
    X=\begin{bmatrix} 0 & 1 \\ 1 & 0 \end{bmatrix}, \ \ \ \ \ \ \begin{matrix} X|0\rangle= |1\rangle \\ X|1\rangle= |0\rangle \end{matrix}
\end{equation}

\begin{equation}\label{eq:hadamard}
    H=\begin{bmatrix} \frac{1}{\sqrt{2}} & \frac{1}{\sqrt{2}} \\ \frac{1}{\sqrt{2}} & -\frac{1}{\sqrt{2}} \end{bmatrix}, \ \ \ \ \ \ \begin{matrix} H|0\rangle=\tfrac{1}{\sqrt{2}}(|0\rangle+|1\rangle)\\ H|1\rangle=\tfrac{1}{\sqrt{2}}(|0\rangle-|1\rangle) \end{matrix}
\end{equation}

\begin{equation}\label{eq:ry}
    R_{y}(\theta)=\begin{pmatrix}
cos\left ( \frac{\theta}{2} \right ) & -sin\left ( \frac{\theta}{2} \right )\\
sin\left ( \frac{\theta}{2} \right ) & cos\left ( \frac{\theta}{2} \right )
\end{pmatrix}
\end{equation}

\begin{gather}\label{eq:cnot}
    CNOT=\begin{bmatrix} 1 & 0 & 0 & 0\\ 0 & 1 & 0 & 0\\ 0 & 0 & 0 & 1\\ 0 & 0 & 1 & 0 \end{bmatrix} \\
    \begin{matrix} CNOT|0\rangle|0\rangle=|0\rangle|0\rangle & CNOT|1\rangle|0\rangle=|1\rangle|1\rangle \\ CNOT|0\rangle|1\rangle=|0\rangle|1\rangle & CNOT|1\rangle|1\rangle=|1\rangle|0\rangle \end{matrix}\nonumber
\end{gather}

\begin{figure}[ht]
    \centering
    \includegraphics[scale=0.85]{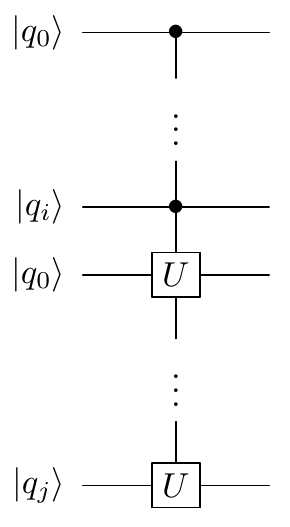}
    \caption{Representation in the quantum circuit of a general controlled gate where $q_0...q_i$ are $i-1$ control qubits and $q_0...q_j$ are target qubits. Operation $U$ is only applied to target qubits if all control qubits have true values.}
    \label{fig:controlled}
\end{figure}

In addition to these, there is a generalization of the CNOT quantum gate, which can include more than one qubit having the control function ($0,...,i$) and more than one qubit as target ($0,...,j$), in addition to being able to apply an arbitrary $U$ operator to the target qubits. The general controlled gate is represented in Figure \ref{fig:controlled}.

\section{Quantum Machine Learning}\label{sec:qml}

As stated in Section~\ref{sec:intro}, several approaches employ the concept of quantum machine learning to perform classification tasks. In this section, we will introduce some of the works most related to our proposal.

In~\cite{blank2020quantum} is presented a binary quantum classifier based on kernel methods. The authors show that quantum computing can improve distance-based classification tasks in machine learning based on kernel methods. This conclusion is obtained through the efficiency with which quantum computing can deal with large feature spaces. In~\cite{park2020theory}, the label qubit is removed reducing the original classifier proposed by~\cite{schuld2017implementing}. They propose a distance-based quantum classifier that uses SWAP-Test to obtain a similarity metric between the training and test samples.

In~\cite{ruan2017quantum} it is proposed a quantum version of the well-known K-Nearest Neighbors (KNN) algorithm. The authors used Hamming distance as a measure of similarity between the test and training samples. To carry out the classification, the features of the dataset are stored in quantum bits. To prove the model's effectiveness, experiments were performed using the MNIST database.

The classifier proposed in~\cite{kathuria2020implementation} also uses the Hamming distance. As in~\cite{blank2020quantum}, SWAP-Test was also used to obtain the similarity measure. The experiments were performed on the real quantum devices \textit{ibmqx2} and \textit{ibmq\_16\_melbourne} from IBM Quantum Experience. Despite demonstrating the applicability and efficiency of the classifier in some examples, the results show that the execution is not yet fully feasible in current quantum devices.

Finally, during the time of writing, we noted the related work~\cite{neumann2020classification}. Despite being a related approach where the authors seek to build a minimal classifier, in~\cite{neumann2020classification} there is still the conditional measurement present in the classifier shown in~\cite{schuld2017implementing}.

\section{Hadamard Classifier}\label{sec:hc}

The Hadamard Classifier~\cite{schuld2017implementing} aims to investigate how to perform a distance-based classification task with a minimal quantum circuit. The strategy used in the HC is to use \textit{amplitude encoding} to encode the input features and perform quantum interference to evaluate the distance from a new input vector to the training (stored) data. HC is a quantum machine learning model that can be implemented in NISQ devices. To validate the HC, the authors performed supervised classification experiments using the Iris dataset~\cite{fisher1936use} (available in Scikit-learn~\cite{scikit-learn}). The quantum system that performs the classification is shown in Eq.~\eqref{eq:hc_state}, where $|m\rangle$ is an index register flagging the $m$th training vector, $|\psi_{\mathbf{x}^m}\rangle$ is the $m$th training vector, $|\psi_{\tilde{\mathbf{x}}}\rangle$ is the new input, and $|y^m\rangle$ is a single qubit that stores the class.

\begin{equation}
|\mathcal{D}\rangle = \frac{1}{\sqrt{2 M}} \sum_{m=1}^M |m\rangle \Big( |0\rangle |\psi_{\tilde{\mathbf{x}}}\rangle + |1\rangle |\psi_{\mathbf{x}^m}\rangle \Big) |y^m\rangle
\label{eq:hc_state}
\end{equation}

\begin{figure}[ht]
    \centering
    \includegraphics[scale=0.78]{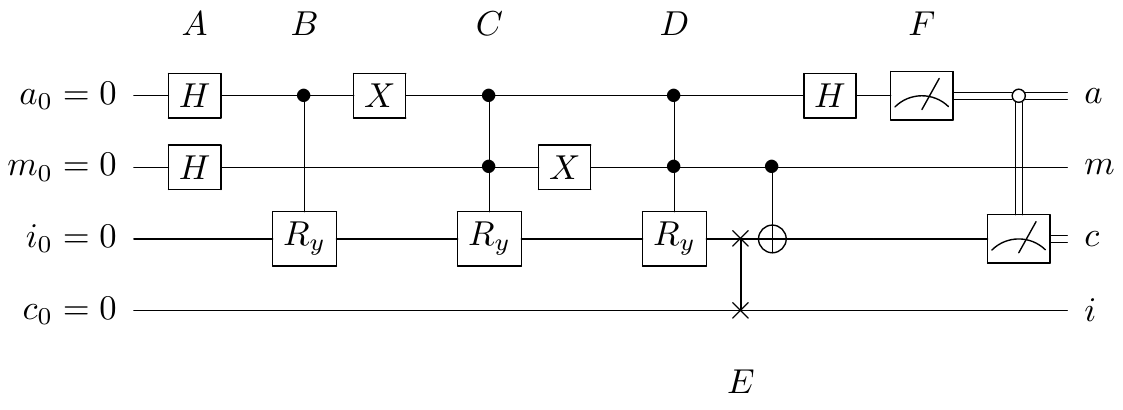}
    \caption{Hadamard Classifier present in~\cite{schuld2017implementing} showing an example of classifier circuit wherein the step B the test vector is loaded and in steps C and D the training vectors are loaded. In step F there is the disentanglement with the Hadamard gate and the 2-qubit measurement.}
    \label{fig:hc}
\end{figure}

The main limitation of HC is a conditional measurement (see Figure~\ref{fig:hc}), called \textit{postselection}, that depends on the probability of measuring $|0\rangle$ in the \textit{ancilla qubit}. A measurement is made in the \textit{class qubit} only after the postselection succeeds. Experimental results showed 100\% accuracy for classes 1 and 2 of the Iris dataset. However, due to the dependence on postselection, more repetitions are needed to obtain a realistic estimate of the result. Also, with the presence of this postselection, it is necessary to perform a 2-qubit measurement, which can increase the probability of errors from the current quantum noisy devices.

\section{Quantum One-class Classifier}\label{sec:qocc}

\begin{figure}[ht]
    \centering
    \includegraphics[scale=0.5]{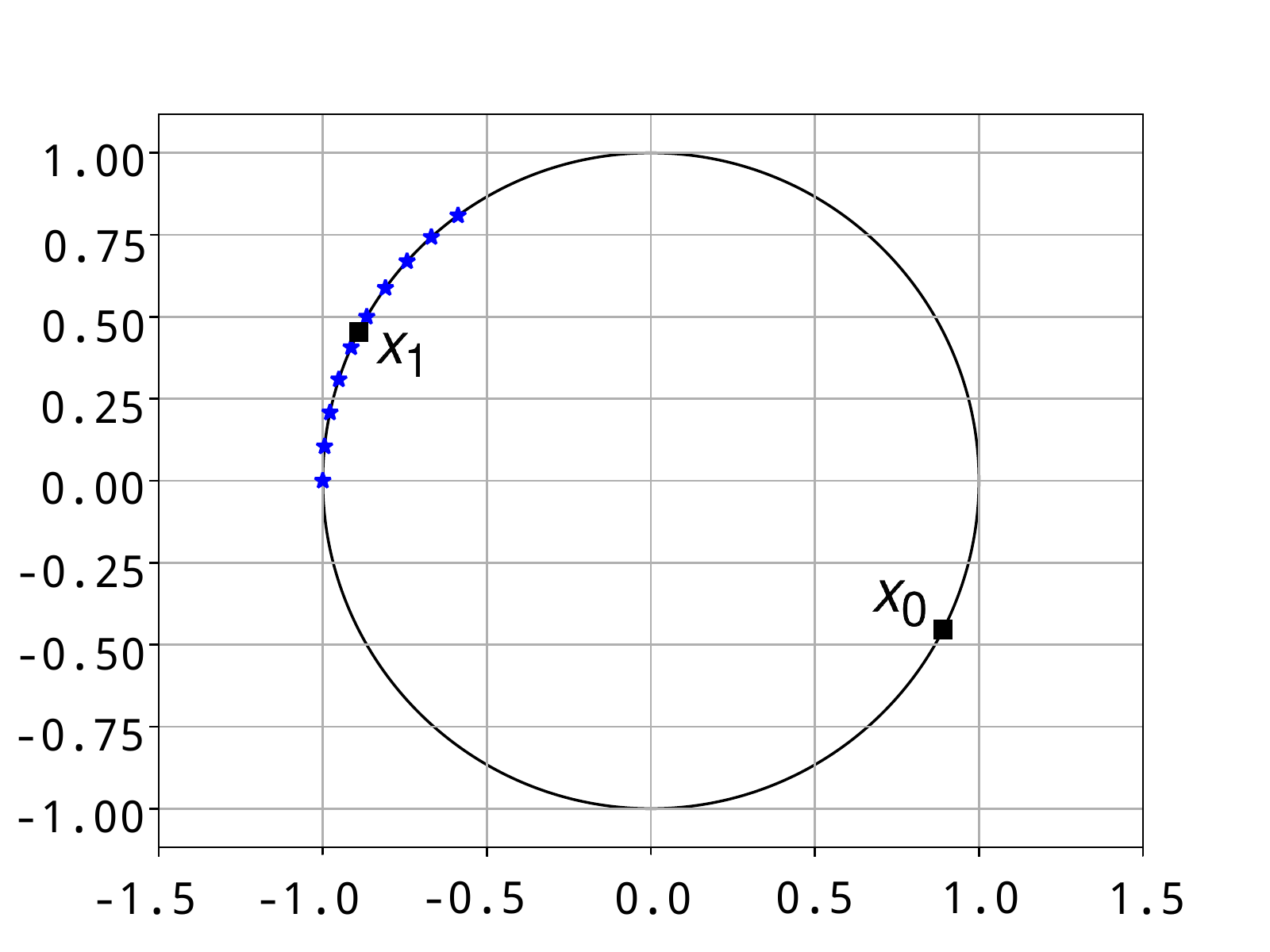}
    \caption{An artificial example where the probability of postselection tends to zero for a given new input vector $\mathbf{x}_0$ from class 0. The stars are samples from the dataset and have a unit norm. The black square $\mathbf{x}_0$ ($\mathbf{x}_1$) is a sample of class 0 (1) to be classified. Postselection is performed based on the distance between the sample to be classified and all other samples in the dataset. In this example, the sample $\mathbf{x}_0$ would not be classified because the approach in~\cite{schuld2017implementing} requires output 0 in the postselection to proceed with the classification. Output 0 in postselection is only achieved if the data to be classified is close to the other data in the dataset.}
    \label{fig:example_pacc}
\end{figure}

The postselection of the HC~\cite{schuld2017implementing} succeeds with probability $p_{acc}=\frac{1}{4M}\sum_m|\tilde{\mathbf{x}} + \mathbf{x}^{m}|^2$. This probability depends on data distribution and can tend to zero. Figure \ref{fig:example_pacc} presents an artificial dataset where the postselection probability is approximately 0.02 for the pattern $\mathbf{x}_0$ and 0.98 for pattern $\mathbf{x}_1$. The postselection probability is a function of the Euclidean distance of the new pattern to the patterns in the dataset and returns 0 with a higher probability if the new pattern is near to the patterns in the dataset. It is important to point out that the post-selection problem found in \cite{schuld2017implementing} has already been addressed in \cite{blank2020quantum}, where it was shown that only the result of measuring a single qubit is necessary to determine the class.

In this Section, we redefine the HC to use the outcome of the postselection as the output of the classifier. With this, we remove the \textit{class qubit}, reduce the number of repetitions necessary to estimate the output of the classifier, and reduce the number of operations necessary to perform the classification.

The QOCC is based on the HC and extends its applicability to allow the classification of a new input vector from a quantum one-class classifier indicating the probability that the vector will be associated with the set of loaded (training) vectors in the classifier. Such a feature allows us to execute the classifier with fewer repetitions to obtain the result and perform only a 1-qubit measurement, instead of the 2-qubit measurement present in the HC. As pointed in~\cite{kim2020quantum}, the error rate present in current quantum devices can cause the power of quantum computing to be hidden. Therefore, the reduction of a 2-qubit measurement to a 1-qubit measurement also means an attempt to mitigate errors from the current noisy quantum computers without necessarily making use of a specific procedure for this.

In our approach, we use the same data preprocessing strategy used in~\cite{schuld2017implementing}. However, the classification of a new input vector works similar to a probabilistic quantum memory~\cite{trugenberger2001probabilistic}. This connection with quantum probabilistic memories comes from the similarity in how the output is defined: a new input is compared with the samples already present in the classifier and the output is given by a probability distribution. Thus, the more distant (different) the new input and stored samples are, the more likely we are to see $|1\rangle$ when measuring the ancilla qubit. On the other hand, if the similarity between them is high, we will see in the output a greater probability of observing $|0\rangle$ in the measurement of the ancilla. Thus, a new input vector is classified according to a \textit{degree of membership} of this new vector against vectors already stored in the classifier. This degree of membership is the probability of measuring 0 on the output where previously was postselection.

Thus, the modification made here allows us to abstract the issue of data distribution, which could lead to a very low probability of postselection. Also, by loading training samples from a single class, our classifier proves to be more flexible by simplifying the classification/association of a new input vector in any class.

\begin{figure}[ht]
    \centering
    \includegraphics[scale=0.85]{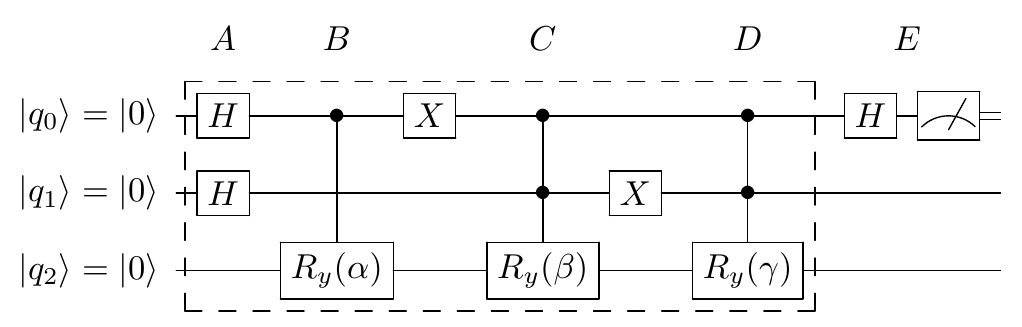}
    \caption{Quantum circuit implementing quantum one-class classifier. The result of successive runs of this circuit represents a kind of degree of membership of the new input vector (step B) to the vectors already stored in the classifier (steps C and D). The quantum gates $R_{y}$ are responsible for loading data from each vector through their associated angles $\alpha$, $\beta$, and $\gamma$. Enclosed the dashed lines is the subcircuit for state preparation. At the end of the computation (step E), the Hadamard gate in the ancilla qubit interferes with the copies of the new input vector with the loaded vectors and then the ancilla is measured.}
    \label{fig:one_class}
\end{figure}

Figure~\ref{fig:one_class} shows the circuit of QOCC receiving inputs with two features. In step E the computed output of successive measurements represents the degree of membership of the new input vector to the $c$ class. Therefore, a degree of membership greater than 0.5 means that the new input vector has been classified as class $c$. The procedure for performing QOCC with 2 stored samples is shown in Algorithm \ref{algo:qoc}. 

\begin{algorithm}[ht]
\caption{Quantum One-class Classifier (QOCC)}
\label{algo:qoc}
\KwIn{$test$, $training$}
Initialize quantum registers $ancilla = |a\rangle$, $index=|m\rangle$, $data=|i\rangle$\\
Perform $H|a\rangle$ and $H|m\rangle$\\
Perform C-$R_{y}(test)|a\rangle|i\rangle$ to load the sample to be classified\\
Apply $X|ancilla\rangle$ to entangle the test sample with the ground state of the ancilla\\
Perform CC-$R_{y}(training[0])|a\rangle|m\rangle|i\rangle$ to load the first training sample\\
Apply $X|m\rangle$ to entangle the first training sample with the ground state of the index and the excited state of the ancilla\\
Perform CC-$R_{y}(training[1])|a\rangle|m\rangle|i\rangle$ to load the second training sample\\
Apply $H|a\rangle$ to interferes the copies of the test sample with the training ones\\
Measure $|a\rangle$ to get the probability of the output being $|0\rangle$
Return the degree of membership accordingly to the result of the measurement
\end{algorithm}

In Figure \ref{fig:one_class} the rotation gates $R_y$ in steps $B$, $C$ and $D$ load the classical input vectors in quantum amplitudes. However, this is not so straightforward and requires an auxiliary procedure to perform such an embedding, to be explained in Section~\ref{subsec:amp_encoding}.

\begin{figure}[ht]
    \centering
    \includegraphics[scale=0.85]{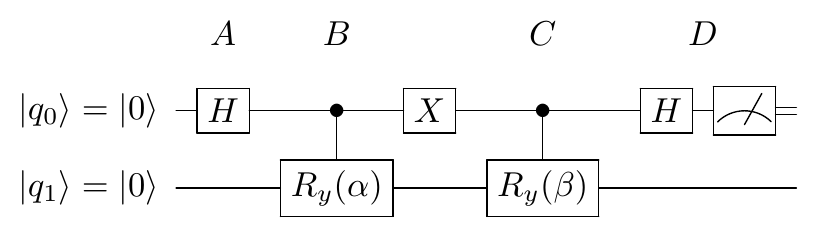}
    \caption{Minimal Quantum One-Class Classification. In step B the test sample is loaded. In step C a single stored sample is loaded. Finally, in step C, the disentanglement and measurement of the ancilla qubit is performed.}
    \label{fig:minimal-qocc}
\end{figure}

Finally, when performing the experiments, we noticed that with only 1 stored (training) sample in the classifier it was possible to obtain an accuracy equivalent to the case where we used 2 training samples. This performance equivalence between the use of 1 or 2 stored samples is justified as this single sample may be close to the centroid representation of the class. In this way, the QOCC just got smaller, eliminating the index qubit that can be seen in Figure~\ref{fig:one_class}. Regarding the Algorithm \ref{algo:qoc}, to perform the classification with 1 stored sample, it is not necessary to have the index qubit. Thus, the QOCC, as a minimum classifier, can be seen in Figure~\ref{fig:minimal-qocc}. The results of these experiments can be seen in Section~\ref{sec:results}.

\subsection{Amplitude Encoding}\label{subsec:amp_encoding}

\begin{figure}[ht]
    \centering
    \includegraphics[scale=0.9]{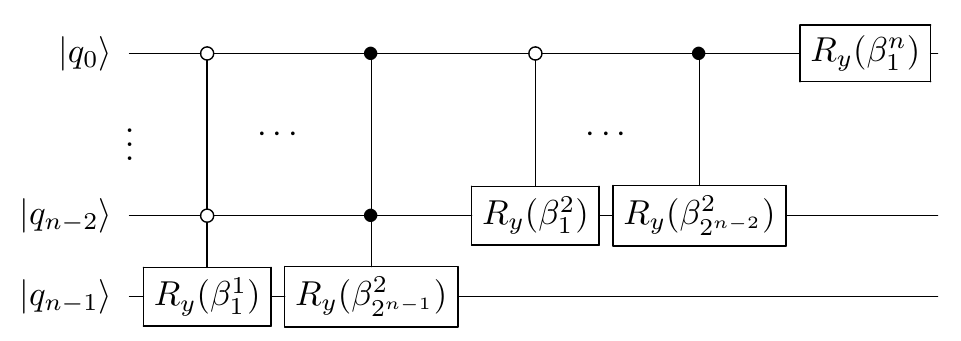}
    \caption{Procedure to reach the vector $|0...0\rangle$ from any other vector.}
    \label{fig:amp_encoding}
\end{figure}

To perform the amplitude encoding necessary to encode input data in the amplitudes of the qubits, we follow the strategy present in~\cite{schuld2018supervised}. Figure~\ref{fig:amp_encoding} illustrates how to start from the desired vector to the state $|0...0\rangle$.

In Figure~\ref{fig:amp_encoding} each $\beta$ is a rotation angle to be performed by the respective $R_{y}$ gate and is calculated according to \eqref{eq:betas}, where $s=\left \{1,...,n\right \}$, $n$ is the number of qubits and $j=\left\{ 1,...,2^{i-1} \right\}$ ($i$ is the index of the qubit in which the rotation gate is being applied).

\begin{equation}\label{eq:betas}
    \beta_{j}^{s}=2\textrm{ arcsin}\left ( \frac{\sqrt{\sum_{l=1}^{2^{s-1}} |\alpha_{(2j-1)2^{s-1}+l}|^{2}}}{\sqrt{\sum_{l=1}^{2^{s}} |\alpha_{(j-1)2^{s}+l}|^{2}}} \right )
\end{equation}

\begin{figure}[ht]
    \centering
    \includegraphics[scale=0.85]{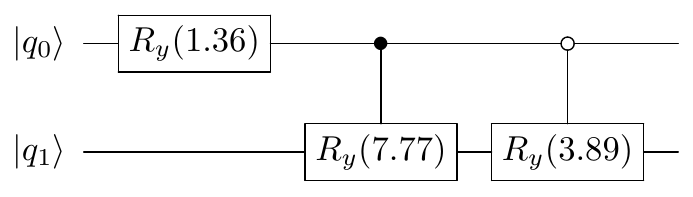}
    \caption{Example of how to load a sample in the amplitudes of a quantum state.}
    \label{fig:amp_encoding_example}
\end{figure}

The main idea is to perform different rotations for each portion of the superposition state through multi-controlled rotation gates. Amplitude coding is performed by applying the inverse gates shown in Figure~\ref{fig:amp_encoding} in the inverse order. As an example of how amplitude encoding works, Figure~\ref{fig:amp_encoding_example} shows the procedure being performed to load the 4 feature input vector $|\psi\rangle = - 0.286|00\rangle + 0.723|01\rangle - 0.464|10\rangle - 0.425|11\rangle$ (Iris sample 20) in a quantum circuit.

\section{Experiments and Results}\label{sec:results}

In this Section, we perform experiments using a quantum simulator and real NISQ quantum devices. As in~\cite{schuld2017implementing}, we use the Iris dataset~\cite{fisher1936use} in our experiments. In addition, we also performed experiments with Haberman's Survival~\cite{haberman1976generalized} and Skin Segmentation~\cite{bhatt2009efficient} datasets. The characteristics of the datasets can be seen in Table~\ref{tab:datasets-characteristics}.

Due to imbalanced data in Haberman's~\cite{haberman1976generalized} and Skin~\cite{bhatt2009efficient} datasets, we use SMOTE re-sample procedure~\cite{chawla2002smote} to increase the number of samples of the minority class. As our goal is to build a binary classification method, we choose the two first classes of the Iris dataset. To use the less number of qubits, achieving the minimal quantum classifier, we chose 2 features of the datasets. In Haberman's, we drop out the attribute that stores the patient's year of operation. The features choose in the Iris dataset are sepal width and petal length. The attributes of the Skin dataset are RGB colors of the pixel, and we chose arbitrarily the R(red) and B(blue) colors. To estimate the performance in an equal number of samples in all datasets, we separate the Haberman's and Skin in 4 batches of the same number of Iris instances of the two first classes (100 samples).

To be able to use amplitude encoding to encode the samples on the quantum circuit the sample vector must be normalized and have unit length. Therefore, we standardize the data features to have zero mean and unit variance and normalize the sample vectors. Just like in HC, an adequate choice of training vectors strongly influences the probability of classification success. Ideally, we should load a larger training set, but due to the limitations and errors of the available quantum hardware, we decided, initially, to keep the original proposal and load only 2 training vectors with 2 features for the classifier - this choice aims to keep the classifier as simple as possible by also configuring it as a proof of concept. Therefore, is called \textit{training stage} the procedure that chooses the pair that achieves better accuracy in a training set. This stage is performed exclusively in the quantum simulator present in Qiskit~\cite{qiskit} on a training set, that corresponds to 70\% of the total dataset. As a way of not having to execute all the combinations present in the training set, we decided to take (among this 70\%) 30 pairs (or samples, in the case of the approach with a single stored sample) to perform the training stage. The remaining 30\% is used as a validation set. To apply a fair statistical comparison, we use the same sample folds on HC and QOCC.

In the validation step, we perform the inferences in the validation set using the fixed pair chosen in the training step. This choice has a different approach depending on the quantum classifier. If the experiment is on QOCC this pair has the same class, and if is on HC has different classes. In addition to simulation using Qiskit simulator, the QOCC, and HC inference is executed on IBM quantum devices~\cite{ibmq} (ibmq\_athens and ibmq\_santiago) followed the default (1024 runs) for each experiment/sample. Due to noise present in real quantum computers, we get the average accuracy over 5 runs of the validation circuits resulting from the Algorithm~\ref{algo:qoc}, with the training pair and a sample from the validation set as input. The complete experiment procedure is present in Algorithm~\ref{algo:exp}.

\begin{table}[!ht]
    \renewcommand{\arraystretch}{1.3}
    \caption{Datasets characteristics}
    \label{tab:datasets-characteristics}
  \centering
  \begin{tabular}{|c|c|c|c|}
    \hline
    \textbf{Dataset} & \textbf{Classes} & \textbf{Features} & \textbf{Instances}\\
    \hline
    Iris & 3 & 4 & 150 \\
    Haberman's Survival & 2 & 3 & 306 \\
    Skin Segmentation & 2 & 3 & 245057 \\
    \hline
  \end{tabular}
\end{table}

\begin{algorithm}[ht]
\SetKwInOut{Input}{input}
\caption{Experiments for Quantum One-Class Classification}
\label{algo:exp}
\Input{Training set $\mathcal{T}$ and validation set $\mathcal{V}$ containing patterns of one class}
Set accuracy variable $a_{best}=0$\\
Set best samples index $k_{best}$\\
Set inference array $I_{inf}^{train}$\\
\For{$(\textbf{x}^k,\textbf{x}^{k+1})\in\mathcal{T}\times \mathcal{T}$ where $k\in\mathbb{N}$}{
    Calculate the angles $\mathcal{P}_{angles}$ of the training pairs for $R_{y}$ gates\\
    $\mathcal{X} = \mathcal{T} - \{\textbf{x}^k,\textbf{x}^{k+1}\}$\\
	\For{$x\in\mathcal{X}$}{
		Calculate test angle $x_{angle}$ for $R_{y}$ gate\\
		$y =$ QOCC$_{simulation}$($x_{angle}$, $\mathcal{P}_{angles}$)\\
		$I_{inf}^{train} \leftarrow (x, y)$\\
	}
	$a_{actual} = accuracy(\mathcal{X},I_{inf}^{train})$\\
	\If{$a_{actual} \geq a_{best}$}{
	   $a_{best}\leftarrow a_{actual}$\\
	   $k_{best}\leftarrow k$\\
	}
	
}
Calculate the angles $\mathcal{P}_{angles}^{best}$ of the training pairs with better accuracy\\ 
\For{$x \in \mathcal{V}$}{
    Calculate test angle $x_{angle}$ for $R_{y}$ gate\\
    $y =$ QOCC$_{real\_device}$($x_{angle}$, $\mathcal{P}_{angles}^{best}$)\\
	$I_{inf}^{val} \leftarrow (x, y)$\\
}
Return $a_{val} = accuracy(\mathcal{V}, I_{inf}^{val})$\\
\end{algorithm}

Therefore, in order to validate our approach, we replicate the HC and compare with our QOCC (Table~\ref{tab:results-new-hc} and Table~\ref{tab:results-new-2}, respectively). For comparison purposes, we perform the same classification tasks using known classical classifiers. The results obtained are shown in Table~\ref{tab:results-classical-2}.

\begin{table}[!t]
    \renewcommand{\arraystretch}{1.3}
    \caption{Results of Hadamard Classifier execution on simulation and on the real devices ibmq\_athens and ibmq\_santiago (marked with an $\ast$).}
    \label{tab:results-new-hc}
  \centering
  \begin{tabular}{|c|c|c|}
    \hline
    \textbf{Dataset} & \textbf{Simul./real} & \textbf{HC} \\
    \hline
    Iris & simul. & 97.78\% \\
    Iris & real & 92,44\% / 92\%$^\ast$ \\
    Haberman's & simul. & 64.72\%\\
    Haberman's & real & 62.50\% / 62.72\%$^\ast$ \\
    Skin & simul. & 95.56\%\\
    Skin & real & 87.56\% / 87.56\%$^\ast$ \\
    \hline
  \end{tabular}
\end{table}

\begin{table}[!t]
    \renewcommand{\arraystretch}{1.3}
    \caption{Results of 2-class QOCC execution with 2 stored samples on simulation and on the real devices ibmq\_athens and ibmq\_santiago (marked with an $\ast$).}
    \label{tab:results-new-2}
  \centering
  \begin{tabular}{|c|c|c|c|}
    \hline
    \textbf{Dataset} & \textbf{Simul./real} & \textbf{QOCC $\mathbf{C_{1}}$} & \textbf{QOCC $\mathbf{C_{2}}$} \\
    \hline
    Iris & simul. & 98.89\% & 100\% \\
    Iris & real & 98.89\% / 91.11\%$^\ast$ & 98.67\% / 97.33\%$^\ast$ \\
    Haberman's & simul. & 65\% & 64.17\% \\
    Haberman's & real & 64.61\% / 63.28\%$^\ast$ & 63.11\% / 63.50\%$^\ast$ \\
    Skin & simul. & 94.17\% & 95.56\% \\
    Skin & real & 89.17\% / 83.44\%$^\ast$ & 90.44\% / 88.44\%$^\ast$ \\
    \hline
  \end{tabular}
\end{table}

\begin{table}[!t]
    \renewcommand{\arraystretch}{1.3}
    \caption{Accuracy of the classical classifiers SVM, DT, KNN, and SGD.}
    \label{tab:results-classical-2}
  \centering
  \begin{tabular}{|c|c|c|c|c|}
    \hline
    \textbf{Dataset} & \textbf{SVM} & \textbf{DT} & \textbf{KNN} & \textbf{SGD} \\
    \hline
    Iris & 100\% & 100\% & 100\% & 100\% \\
    Haberman's & 65\% & 65.83\% & 73.33\% & 58.33\% \\
    Skin & 95.56\% & 95.83\% & 95.56\% & 86.11\% \\
    \hline
  \end{tabular}
\end{table}

As noted in Section~\ref{sec:qocc}, in addition to removing the class qubit, it was also possible to obtain significant results with competitive accuracy when running QOCC with only 1 stored sample. With that, it was possible to remove the index qubit and improve QOCC in such a way as to be a minimum model of classification based on distance. The results for these experiments can be seen in Table~\ref{tab:results-qocc-1-sample}. Regarding Algorithms~\ref{algo:qoc} and \ref{algo:exp}, using only 1 stored sample, instead of loading pairs of samples, we only load 1 training/validation sample. Lastly, the repository https://github.com/lucasponteslpa/QOCClassifier contains all codes used to generate the results of this paper.

\begin{table}[!t]
    \renewcommand{\arraystretch}{1.3}
    \caption{Results of 2-class QOCC execution with 1 stored sample on simulation and on the real devices ibmq\_athens and ibmq\_santiago (marked with an $\ast$).}
    \label{tab:results-qocc-1-sample}
  \centering
  \begin{tabular}{|c|c|c|c|}
    \hline
    \textbf{Dataset} & \textbf{Simul./real} & \textbf{QOCC $\mathbf{C_{1}}$} & \textbf{QOCC $\mathbf{C_{2}}$} \\
    \hline
    Iris & simulation & 98.89\% & 98.89\% \\
    Iris & real & 97.78\% / 98.89\%$^\ast$ & 98.89\% / 98.89\%$^\ast$ \\
    Haberman's & simulation & 62.78\% & 64.17\% \\
    Haberman's & real & 63.33\% / 61.94\%$^\ast$ & 64.44\% / 61.11\%$^\ast$ \\
    Skin & simulation & 95.56\% & 96.67\% \\
    Skin & real & 95\% / 95.83\%$^\ast$ & 96.11\% / 96.39\%$^\ast$ \\
    \hline
  \end{tabular}
\end{table}

\subsection{Discussion}

The accuracy presented in \cite{schuld2017implementing} is only achieved if postselection succeeds (i.e. when ancilla qubit is 0). In the experiments performed, postselection had a probability greater than 50\% of success in only 47.66\% of the executions on the real quantum computer. This shows that HC requires more executions to be carried out to have a more accurate answer.

Regarding the impossibility of classifying vectors from the class 2 pointed in~\cite{schuld2017implementing}: this obstacle was based on the rapid decoherence of the class qubit when storing state $|1\rangle$. In our experiments, it was not observed this impediment in classifying class 2 vectors. 

As our goal is to build a minimum distance-based classifier, we ran our final experiments with just 1 stored sample. The results of such experiments are shown in Table~\ref{tab:results-qocc-1-sample} and show that the accuracy, remained consistent with the experiments containing 2 stored samples (Table~\ref{tab:results-new-2}). We can also observe that, for the Skin dataset, QOCC with 1 stored sample performed better than that with 2 stored samples. The advantage of QOCC over HC at this point may not seem significant, since we are dealing with only 1 class at a time. However, such a reduction becomes more noticeable in a multiclass approach with multiple QOCCs running in parallel. Thus, the execution of a multiclass classifier based on QOCC on NISQ devices would become more feasible concerning the number of qubits needed.

Like in HC, our QOCC was built to deal only with the real part of quantum states and is not suitable for general quantum states. An approach to quantum classification that also takes into account the imaginary part of quantum states is the one that uses the SWAP-test, as in \cite{blank2020quantum}.

Finally, it is necessary to note that on datasets with only two classes the QOCC works as a complete quantum binary classifier. Therefore, it is possible to classify the entire dataset only concerning the samples of a single class stored in the classifier, so that the high (low) degree of membership of the test samples to those already stored gives us the classification between the two classes.

\section{Conclusion}
\label{sec:conclusion}
In this work, we proposed a minimal Quantum One-Class Classifier based on the Hadamard Classifier and the idea of probabilistic quantum memory. Compared with the HC, QOCC has equivalent or improved accuracy on real quantum devices and uses fewer quantum resources. The QOCC is competitive when compared with classical classifiers and can be used to classify quantum data.

QOCC shows an advantage regarding the HC concerning its size and the reduction of the 2-qubit measurement to a 1-qubit measurement. In datasets with two classes, the QOCC behaves like a complete quantum binary classifier, indicating in the ancilla qubit the degree of membership of the input vector to the class, just needing to load data from one of the classes. We also provide an update of the HC performance in current quantum devices.

There are some possible improvements to be explored. Possible future works are to investigate how to use the QOCC to classify multiclass datasets and explore how to insert parameters into the classifier to improve its accuracy. Also, it is possible to conduct a circuit optimization study for even more efficient execution on NISQ devices.

\section*{Acknowledgments}
This work was supported by CNPq, CAPES, and FACEPE (Brazilian research agencies). We acknowledge the use of IBM Quantum services for this work. The views expressed are those of the authors and do not reflect the official policy or position of IBM or the IBM Quantum team.


\begin{thebibliography}{10}
\providecommand{\url}[1]{#1}
\csname url@samestyle\endcsname
\providecommand{\newblock}{\relax}
\providecommand{\bibinfo}[2]{#2}
\providecommand{\BIBentrySTDinterwordspacing}{\spaceskip=0pt\relax}
\providecommand{\BIBentryALTinterwordstretchfactor}{4}
\providecommand{\BIBentryALTinterwordspacing}{\spaceskip=\fontdimen2\font plus
\BIBentryALTinterwordstretchfactor\fontdimen3\font minus
  \fontdimen4\font\relax}
\providecommand{\BIBforeignlanguage}[2]{{%
\expandafter\ifx\csname l@#1\endcsname\relax
\else
\language=\csname l@#1\endcsname
\fi
#2}}
\providecommand{\BIBdecl}{\relax}
\BIBdecl

\bibitem{nielsen2010quantum}
M.~A. Nielsen and I.~L. Chuang, \emph{Quantum Computation and Quantum
  Information}.\hskip 1em plus 0.5em minus 0.4em\relax Cambridge University
  Press, 2010.

\bibitem{shor1994}
\BIBentryALTinterwordspacing
P.~W. Shor, ``Algorithms for quantum computation: Discrete logarithms and
  factoring,'' in \emph{Proceedings of the 35th Annual Symposium on Foundations
  of Computer Science}, ser. SFCS '94.\hskip 1em plus 0.5em minus 0.4em\relax
  Washington, DC, USA: IEEE Computer Society, 1994, pp. 124--134. [Online].
  Available: \url{http://dx.doi.org/10.1109/SFCS.1994.365700}
\BIBentrySTDinterwordspacing

\bibitem{grover1996}
\BIBentryALTinterwordspacing
L.~K. Grover, ``A fast quantum mechanical algorithm for database search,'' in
  \emph{Proceedings of the Twenty-eighth Annual ACM Symposium on Theory of
  Computing}, ser. STOC '96.\hskip 1em plus 0.5em minus 0.4em\relax New York,
  NY, USA: ACM, 1996, pp. 212--219. [Online]. Available:
  \url{http://doi.acm.org/10.1145/237814.237866}
\BIBentrySTDinterwordspacing

\bibitem{biamonte2017quantum}
J.~Biamonte, P.~Wittek, N.~Pancotti, P.~Rebentrost, N.~Wiebe, and S.~Lloyd,
  ``Quantum machine learning,'' \emph{Nature}, vol. 549, no. 7671, pp.
  195--202, 2017.

\bibitem{blank2020quantum}
C.~Blank, D.~K. Park, J.-K.~K. Rhee, and F.~Petruccione, ``Quantum classifier
  with tailored quantum kernel,'' \emph{npj Quantum Information}, vol.~6,
  no.~1, pp. 1--7, 2020.

\bibitem{park2020theory}
D.~K. Park, C.~Blank, and F.~Petruccione, ``The theory of the quantum
  kernel-based binary classifier,'' \emph{Physics Letters A}, p. 126422, 2020.

\bibitem{kathuria2020implementation}
K.~Kathuria, A.~Ratan, M.~McConnell, and S.~Bekiranov, ``Implementation of a
  hamming distance--like genomic quantum classifier using inner products on
  ibmqx2 and ibmq\_16\_melbourne,'' \emph{Quantum Machine Intelligence},
  vol.~2, no.~1, pp. 1--26, 2020.

\bibitem{ruan2017quantum}
Y.~Ruan, X.~Xue, H.~Liu, J.~Tan, and X.~Li, ``Quantum algorithm for k-nearest
  neighbors classification based on the metric of hamming distance,''
  \emph{International Journal of Theoretical Physics}, vol.~56, no.~11, pp.
  3496--3507, 2017.

\bibitem{schuld2017implementing}
\BIBentryALTinterwordspacing
M.~Schuld, M.~Fingerhuth, and F.~Petruccione, ``Implementing a distance-based
  classifier with a quantum interference circuit,'' \emph{EPL}, vol. 119,
  no.~6, p. 60002, 2017. [Online]. Available:
  \url{https://doi.org/10.1209/0295-5075/119/60002}
\BIBentrySTDinterwordspacing

\bibitem{neumann2020classification}
N.~M. Neumann, ``Classification using a two-qubit quantum chip,'' \emph{arXiv
  preprint arXiv:2004.10426}, 2020.

\bibitem{Preskill2018quantumcomputingin}
\BIBentryALTinterwordspacing
J.~Preskill, ``Quantum {C}omputing in the {NISQ} era and beyond,''
  \emph{{Quantum}}, vol.~2, p.~79, Aug. 2018. [Online]. Available:
  \url{https://doi.org/10.22331/q-2018-08-06-79}
\BIBentrySTDinterwordspacing

\bibitem{trugenberger2001probabilistic}
C.~A. Trugenberger, ``Probabilistic quantum memories,'' \emph{Physical Review
  Letters}, vol.~87, no.~6, p. 067901, 2001.

\bibitem{ibmq}
\BIBentryALTinterwordspacing
(2020) {IBM} {Q} {E}xperience. [Online]. Available:
  \url{https://www.ibm.com/quantum-computing/technology/experience/}
\BIBentrySTDinterwordspacing

\bibitem{fisher1936use}
R.~A. Fisher, ``The use of multiple measurements in taxonomic problems,''
  \emph{Annals of eugenics}, vol.~7, no.~2, pp. 179--188, 1936.

\bibitem{scikit-learn}
F.~Pedregosa, G.~Varoquaux, A.~Gramfort, V.~Michel, B.~Thirion, O.~Grisel,
  M.~Blondel, P.~Prettenhofer, R.~Weiss, V.~Dubourg, J.~Vanderplas, A.~Passos,
  D.~Cournapeau, M.~Brucher, M.~Perrot, and E.~Duchesnay, ``Scikit-learn:
  Machine learning in {P}ython,'' \emph{Journal of Machine Learning Research},
  vol.~12, pp. 2825--2830, 2011.

\bibitem{kim2020quantum}
C.~Kim, K.~D. Park, and J.-K. Rhee, ``Quantum error mitigation with artificial
  neural network,'' \emph{IEEE Access}, vol.~8, pp. 188\,853--188\,860, 2020.

\bibitem{schuld2018supervised}
M.~Schuld and F.~Petruccione, \emph{Supervised learning with quantum
  computers}.\hskip 1em plus 0.5em minus 0.4em\relax Springer, 2018, vol.~17.

\bibitem{haberman1976generalized}
S.~J. Haberman, ``Generalized residuals for log-linear models,'' in
  \emph{Proceedings of the 9th international biometrics conference}, 1976, pp.
  104--122.

\bibitem{bhatt2009efficient}
R.~B. Bhatt, G.~Sharma, A.~Dhall, and S.~Chaudhury, ``Efficient skin region
  segmentation using low complexity fuzzy decision tree model,'' in \emph{2009
  Annual IEEE India Conference}.\hskip 1em plus 0.5em minus 0.4em\relax IEEE,
  2009, pp. 1--4.

\bibitem{chawla2002smote}
N.~V. Chawla, K.~W. Bowyer, L.~O. Hall, and W.~P. Kegelmeyer, ``Smote:
  synthetic minority over-sampling technique,'' \emph{Journal of artificial
  intelligence research}, vol.~16, pp. 321--357, 2002.

\bibitem{qiskit}
Qiskit, https://qiskit.org/, last accessed: 2019-11-14.

\end{thebibliography}
\end{document}